\documentclass[amsmath,amssymb,twocolumn,aps,pra,nobibnotes]{revtex4-1}
\usepackage{braket,hyperref,graphicx}
\pdfoutput=1
\newcommand{\abs}[1]{\left\lvert{#1}\right\rvert}
\newcommand{\prj}[1]{{\Ket{#1}\Bra{#1}}}
\newcommand{\Hb}{\mathcal{H}}
\DeclareMathOperator{\tr}{tr}
\DeclareMathOperator{\rank}{rank}

\hypersetup{pdfauthor={Matthew F. Pusey and Terry Rudolph},pdftitle={quantum lost property: an operational meaning for the hilbert-schmidt product?}}
\begin{document}
\title{Quantum lost property: A possible operational meaning for the Hilbert-Schmidt product}
\date{October 15, 2012}

\author{Matthew F. Pusey}
\email{m@physics.org}
\author{Terry Rudolph}
\affiliation{Department of Physics, Imperial College London, Prince Consort Road, London SW7 2AZ, United Kingdom}
\begin{abstract}
  Minimum error state discrimination between two mixed states $\rho$ and $\sigma$ can be aided by the receipt of ``classical side information'' specifying which states from some convex decompositions of $\rho$ and $\sigma$ apply in each run. We quantify this phenomena by the average trace distance, and give lower and upper bounds on this quantity as functions of $\rho$ and $\sigma$. The lower bound is simply the trace distance between $\rho$ and $\sigma$, trivially seen to be tight. The upper bound is $\sqrt{1 - \tr(\rho\sigma)}$, and we conjecture that this is also tight. We reformulate this conjecture in terms of the existence of a pair of ``unbiased decompositions'', which may be of independent interest, and prove it for a few special cases. Finally, we point towards a link with a notion of non-classicality known as preparation contextuality.
\end{abstract}
\maketitle

Suppose a system has been prepared in one of two non-orthogonal quantum states. The task of measuring the system in order to estimate which state was used is known as state discrimination \cite{sd1,sd2}, an important concept in quantum information theory. The impossibly of succeeding at this task with certainty enables quantum cryptography \cite{crypto}. Here we investigate a version of state discrimination where, in each run, additional classical information about each of the possible preparations is provided to the agent attempting the discrimination.

\textit{Classical analogy.} Charlie spots the dim outline of a pencil case under his desk. He knows Alice and Bob have both recently lost theirs, and judges the case equally likely to belong to either of them. All the pencil cases at his school are either pink or blue. Charlie believes that girls buy pink pencil cases with probability $1/2$ whilst boys buy them with probability $1/4$. He therefore resolves to return the pencil case to Alice if it is pink, and Bob if it is blue. He calculates the probability of returning the case to its true owner as $\left(1 + \delta_C\right)/2$, where
\begin{equation}
  \delta_C(\{p_i\}, \{q_i\}) = \frac12 \sum_i \abs{p_i - q_i}
\end{equation}
is here equal to $1/4$. Unsatisfied, he devises a better plan: he will ask Alice and Bob what colour their pencil cases actually are, returning it to whoever states the correct colour. The only way this strategy can fail is if Alice and Bob happen to have bought the same colour, in which case Charlie is forced to toss a coin. Hence his probability of success is slightly better, $\left(1 + P_\text{diff}\right)/2$ where
\begin{equation}
  P_\text{diff}(\{p_i\}, \{q_i\}) = 1 - \sum_i p_i q_i
\end{equation}
is $1/2$ in this case.

\textit{Definitions.} Fix a finite dimensional Hilbert space $\Hb$. The optimum probability of discriminating two states $\rho, \sigma \in L(\Hb)$ (with equal priors) is $\left(1 + \delta\right)/2$, where the quantum trace distance $\delta$ is given by \cite{nc}
\begin{equation}
  \delta(\rho, \sigma) = \frac12 \tr\abs{\rho - \sigma}.
\end{equation}
Decomposing $\rho = \sum_i p_i \rho_i$ and $\sigma = \sum_j q_j \sigma_j$ ($p_i, q_j>0$, $\rho_i, \sigma_j$ states), we can define the average trace distance
\begin{equation}
  \Delta(\{p_i\},\{\rho_i\}, \{q_j\}, \{\sigma_j\}) = \sum_{i,j} p_i q_j \delta(\rho_i, \sigma_j).
\end{equation}
If, when attempting to distinguish two states $\rho$ and $\sigma$, we are told in each run which (independently sampled) $\rho_i$ and $\sigma_j$ applies, the best strategy is clearly to optimally distinguish $\rho_i$ from $\sigma_j$. The overall probability of success will then be $\left(1 + \Delta\right)/2$. $\Delta$ was briefly mentioned in Ref.~\cite{dsens}, but a different quantity $D^K$ where the product distribution $p_iq_j$ is replaced by an adversely correlated distribution was deemed preferable in that setting.

\textit{Lower bound.} By the joint convexity \cite{nc} of $\delta$, we have
\begin{equation}
  \Delta(\{p_i\},\{\rho_i\}, \{q_j\}, \{\sigma_j\}) \geq \delta(\rho, \sigma).\label{lower}
\end{equation}
This bound is saturated by the trivial decomposition $p_1 = q_1 = 1$, $\rho_1 = \rho$, $\sigma_1 = \sigma$.

\textit{Upper bound.} By Eq.~\eqref{lower} a decomposition that maximizes $\Delta$ can always be taken to consist of pure states $\rho_i = \prj{\psi_i}$ and $\sigma_j = \prj{\phi_j}$, and so we consider only this case from now on. Hence \cite{nc} $\delta(\rho_i, \rho_j) = \sqrt{1 - \abs{\Braket{\psi_i | \phi_j}}^2} = \sqrt{1 - \tr(\rho_i\sigma_j)}$. Noting that $\sqrt{1 - x}$ is concave \cite{convex} on its domain $x \leq 1$, the trace is linear, and $\sum_{i,j} p_i q_j \rho_i \sigma_j = \rho\sigma$, we have
\begin{equation}
  \Delta = \sum_{i,j} p_iq_j\sqrt{1 - \tr(\rho_i\sigma_j)} \leq \sqrt{1 - \tr(\rho\sigma)}\label{upper}.
\end{equation}

\textit{Saturating the upper bound.} Since $\sqrt{1-x}$ is in fact strictly concave, equality in Eq.~\eqref{upper} can only be achieved if the arguments $x$ in each term of sum (except those with zero probability, which we can remove from the decompositions) are equal. Hence the upper bound is tight for a particular $\rho$ and $\sigma$ if and only if there exists decompositions $\rho = \sum_i p_i \prj{\psi_i}$ and $\sigma = \sum_j q_j \prj{\phi_j}$ which are ``unbiased'' in that $\abs{\Braket{\psi_i | \phi_j}}^2 = \tr{\rho\sigma}$. (Note that by the linearity of the trace, \emph{any} decompositions satisfy the weaker condition $\sum_{i,j} p_i q_j \abs{\Braket{\psi_i | \phi_j}}^2 = \tr(\rho\sigma)$.)

Since numerics indicate that Eq.~\eqref{upper} is tight, we conjecture that a pair of unbiased decompositions exists for any pair of states $\rho$ and $\sigma$. We also make the stronger conjecture that such a pair exists with both decompositions minimal, i.e. $i \in \{1, \dotsc, \rank(\rho)\}, j \in \{1, \dotsc, \rank(\sigma)\}$. We will now prove some special cases of this conjecture.

\begin{figure}
  \begin{center}
    \includegraphics{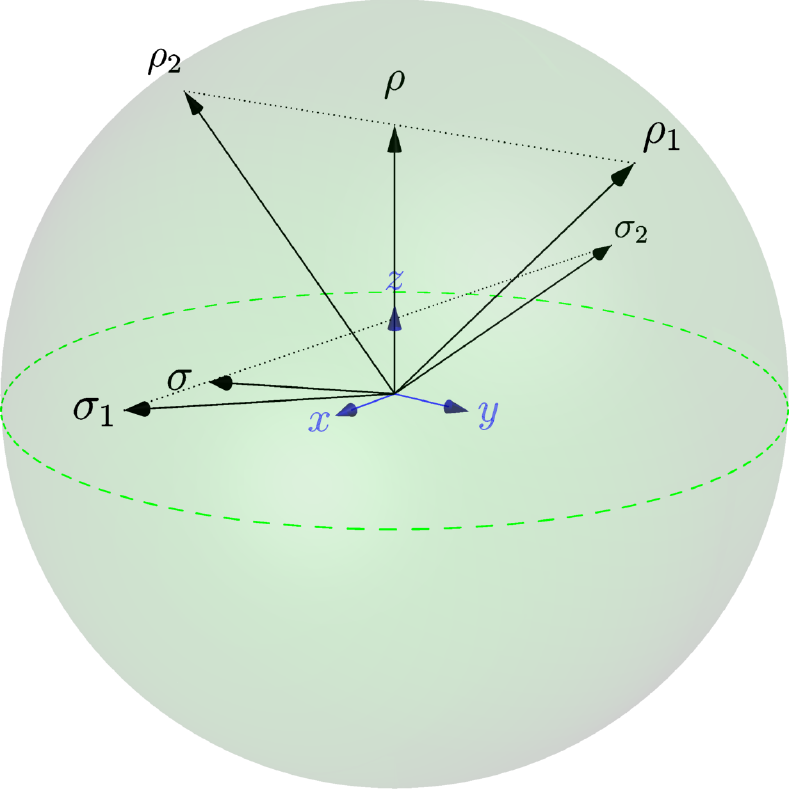}
  \end{center}
  \caption{A pair of unbiased decompositions.}\label{qdecomp}
\end{figure}

\textit{Qubits.} Suppose $\dim \Hb = 2$. Choose a basis so that the Bloch vectors for $\rho$ and $\sigma$ are $\vec\rho = (0,0,r)$ and $\vec\sigma = (s_x, 0, s_z)$ respectively. Then $\vec \rho$ is clearly on the line between the two pure states at $\vec \rho_{1,2} = (0,\pm\sqrt{1 - r^2}, r)$, giving rise to a valid decomposition, and similarly for $\vec \sigma_{1,2} = (\pm\sqrt{1 - s_z^2}, 0, s_z)$. Finally $\vec \rho_i \cdot \vec \sigma_j = rs_z = \vec \rho \cdot \vec \sigma$ and so $\tr(\rho_i \sigma_j) = \tr(\rho\sigma)$ as required. These decompositions are illustrated in Fig.~\ref{qdecomp}.

\textit{Maximally mixed $\sigma$.} Suppose $\dim \Hb = d$, and $\sigma = I/d$. Choose a basis $\{\ket{\psi_i}\}$ in which $\rho$ is diagonal. Clearly there exists a decomposition using these states. Let $\{\ket{\phi_j}\}$ form a basis that is mutually unbiased with respect to the $\{\ket{\psi_i}\}$ basis, for example by using the quantum Fourier transform unitary \cite{nc}. We have that $\sigma = \sum_j q_j \prj{\phi_j}$ with $q_j = 1/d$ and the decompositions are, by construction, unbiased.

\textit{A useful lemma.} Let $f$ be a convex-linear map from the set of states on $\Hb$ to the real numbers. Then any state $\rho$ has a decomposition into $\rank(\rho)$ pure states $\rho_i$ which all satisfy $f(\rho_i) = f(\rho)$.

The proof is as follows. For an arbitrary minimal decomposition $\{\rho_i\}$, consider the figure of merit
\begin{equation}
  F = \sum_i \abs{f(\rho_i) - f(\rho)}
\end{equation}
If $F > 0$ we can construct a new decomposition with smaller $F$ as follows. Take $k$ so that $f(\rho_k)$ is maximal and $l$ so that $f(\rho_l)$ is minimal. Notice that we can ``continuously swap'' $\rho_k$ and $\rho_l$. More formally, there exists continuous functions $\rho_k(\theta), \rho_l(\theta)$ with $\rho_k(0) = \rho_l(\pi) = \rho_k$ and $\rho_k(\pi) = \rho_l(0) = \rho_l$ such that $\rho$ can be decomposed into $\rho_k(\theta), \rho_l(\theta)$ and the $\rho_i$ with $i \neq k,l$ for any $\theta \in [0,\pi]$. To see this, consider the continuous family of unitaries $U(\theta)$ with $U(\theta)\ket{k} = \cos(\theta/2)\ket{k} - \sin(\theta/2)\ket{l}$ and $U(\theta)\ket{l} = \sin(\theta/2)\ket{k} + \cos(\theta/2)\ket{l}$ and all other $\ket{i}$ unaffected, and apply Schr\"odinger's mixture theorem \cite{schr,HJW}. Now by the intermediate value theorem there exists a $\theta^* \in (0,\pi)$ with $f\left( \rho_k(\theta^*)\right) = f\left( \rho_l(\theta^*) \right)$. Since by convex-linearity the average value of $f$ of this new decomposition must still equal $f(\rho)$, this procedure must have reduced $F$. Finally, since the unitary group is compact, the set of decompositions of $\rho$ into pure states is compact and hence $F=0$ must be achieved for some decomposition.

\textit{Corollary: unbiased decomposition of $\rho$.} If $\rho_i$ and $\sigma_j$ are unbiased decompositions, then by linearity
\begin{equation}
  \tr(\rho_i \sigma) = \sum_j q_j \tr(\rho_i \sigma_j) = \sum_j q_j\tr(\rho\sigma) = \tr(\rho\sigma).
\end{equation}
Conversely, setting $f(\cdot) = \tr(\cdot \sigma)$ in the above lemma implies that there always exists a minimal decomposition of $\rho$ satisfying $\tr(\rho_i \sigma) = \tr(\rho\sigma)$. Notice that the proof of the lemma suggests a numerical method for finding such decompositions using a series of one-dimensional search problems, which may sometimes be faster than solving the direct $(d^2-1)$-dimensional search problem.

\textit{Pure $\sigma$.} Suppose that $\rank(\sigma) = 1$. By the above corollary we can decompose $\rho$ into pure states $\rho_i$ such that $\tr(\rho_i \sigma) = \tr(\rho \sigma)$. Since $\sigma$ is already pure we can take $\sigma_1 = \sigma$ and we have a pair of unbiased decompositions.

\textit{Rank two $\sigma$.} Suppose $\rank(\sigma) = 2$. If $\rank(\rho) = 1$ then we are in the previous case, so assume $\rank(\rho) \geq 2$. By the above corollary we can decompose $\sigma$ into two states $\sigma_j = \prj{\phi_j}$ satisfying $\tr(\rho\sigma_j) = \tr(\rho\sigma)$. Apply the corollary again to obtain a decomposition $\rho_i' = \prj{\psi_i'}$ of $\rho$ satisfying $\abs{\Braket{\psi_i'|\phi_1}}^2 = \tr(\rho_i' \sigma_1) = \tr(\rho\sigma_1) = \tr(\rho\sigma)$.

Choose a basis $\ket{1},\cdots,\ket{n}$ ($n = \rank(\rho) \geq 2$) for the support of $\rho$ such that $\ket{2},\cdots\ket{n}$ are orthogonal to $\ket{\phi_1}$. Then $\ket{\psi_i'}$ must be of the form $\sum_k c_k \ket{k}$ where $\abs{c_1} = \sqrt{\tr(\rho\sigma)}/\abs{\Braket{1|\phi_1}}$. Furthermore any state $\ket{\psi}$ of this form also satisfies $\abs{\braket{\psi|\phi_1}} = \tr(\rho\sigma)$ and such states form a connected set. Since $\sum_i p_i \tr(\rho_i' \sigma_2) = \tr(\rho\sigma_2) = \tr(\rho\sigma)$ there must be a $k$ with $\tr(\rho_k'\sigma_2) \geq \tr(\rho\sigma)$ and an $l$ with $\tr(\rho_l'\sigma_2) \leq \tr(\rho\sigma)$. By the above observations and the intermediate value theorem, there is a state $\ket{\psi_1}$ in the support of $\rho$ with $\abs{\Braket{\psi_1|\phi_2}}^2 = \tr(\rho\sigma)$.

Let $p_1$ be maximal, i.e. $p_1 = 1/\Braket{\psi_1 | \rho^{-1} | \psi_1}$ \cite{nc}. $\rho' = (\rho - p_1\prj{\psi_1})/(1 - p_1)$ then has $\rank(\rho') = n-1$ and also satisfies $\tr(\rho'\sigma_j) = \tr(\rho\sigma)$. If $\rho'$ is pure then take it as $\rho_2$ and we are done, otherwise iterate the above procedure to obtain $\ket{\psi_2}$, and so on.

Numerics (using \cite{suparam}) indicate that, when $\rank(\sigma) > 2$, if one simply takes any decomposition $\sigma_j$ with $\tr(\rho \sigma_j) = \tr(\rho\sigma)$ then it is not always possible to find a decomposition of $\rho$ which is unbiased with respect to that $\sigma_j$. This would prevent the above being extended to general $\sigma$.

\textit{Preparation contextuality.} Consider the special case $\rho = \sigma = I/d$. We have shown that one can find two minimal decompositions of $\rho$ with $\Delta = \sqrt{1 - \tr(\rho^2)} = \sqrt{1 - 1/d}$. If, as suggested by the fact they give rise to the same mixed state, there is no actual difference between these two decompositions, it is somewhat surprising that this is larger than the value we obtain if we instead use two identical minimal decompositions of $\rho$, easily seen to be $\Delta = 1 - 1/d$.

This can be made precise by supposing that the two decompositions were represented by a preparation noncontextual ontological model \cite{context}. Briefly, this associates each state $\rho$ with a probability distribution $\mu_\rho(\lambda)$ over ``ontic states'' $\lambda$ (representing the physical state of affairs). Preparation noncontextuality is the assumption that this distribution depends only on $\rho$. Each ontic state $\lambda$ and measurement procedure $M$ gives rise to a probability distribution $p(k|M,\lambda)$ over outcomes $k$, and the quantum statistics are recovered as $p(k|M,\rho) = \int p(k|M,\lambda)\mu_\rho(\lambda)d\lambda$. It is not difficult to see that if some measurement procedure $M$ distinguishes $\rho$ and $\sigma$ with probability $(1+\delta)/2$ then $\mu_\rho$ and $\mu_\sigma$ must be distinguishable with probability at least $(1+\delta_C)/2$, and so every for every $\rho$ and $\sigma$, $\delta_C(\mu_\rho, \mu_\sigma) \geq \delta(\rho, \sigma)$.

If $\sum_i p_i \rho_i$ and $\sum_j q_j \sigma_j$ are minimal decompositions of $I/d$ then we must have $p_i = q_j = 1/d$ and in the model
\begin{equation}
  \mu_{I/d} = \frac{1}d\sum_i \mu_{\rho_i} = \frac1d \sum_j \mu_{\sigma_j}.
  \label{nonctx}
\end{equation}
Since, as argued above, $\delta \leq \delta_C$, we must have
\begin{equation}
  \Delta \leq \Delta_C = \frac{1}{d^2}\sum_{i,j} \delta_C\left( \mu_{\rho_i}, \mu_{\sigma_j} \right).
\end{equation}
By considering the regions where $\mu_0 < \mu_1$ and $\mu_0 \geq \mu_1$ separately and using normalization it can be shown that $\delta_C(\mu_0, \mu_1) = 1 - \int \min\left( \mu_0(\lambda), \mu_1(\lambda) \right)d\lambda$. Hence
\begin{equation}
  \Delta \leq 1 - \frac1{d^2} \int \sum_{i,j} \min\left( \mu_{\rho_i}(\lambda), \mu_{\sigma_j}(\lambda) \right) d\lambda.
\end{equation}
Notice that for any $j$ and $\lambda$, $\sum_i \min\left( \mu_{\rho_i}(\lambda), \mu_{\sigma_j}(\lambda) \right)$ either contains at least one $\mu_{\sigma_j}$, or is equal to $\sum_i \mu_{\rho_i}$ which is equal to $\sum_k \mu_{\sigma_k}$ by Eq.~\eqref{nonctx}. Either way, it is greater than or equal to $\mu_{\sigma_j}$ and so
\begin{equation}
  \Delta \leq 1 - \frac1{d^2} \int \sum_{j} \mu_{\sigma_j} d\lambda = 1 - \frac1d,
\end{equation}
where the equality is by the normalization of the $\mu_{\sigma_j}$. This is indeed exactly the value we get by using two identical decompositions $\rho_i = \sigma_i$, and so any protocol that has a higher probability of success (for example our optimal one) is a proof if preparation contextuality.

\textit{Conclusions.} The fact that mixed states have many decompositions into pure states is a key feature of quantum mechanics, sometimes considered the definition of non-classicality \cite{jon}. We have discussed a task that puts this feature centre stage. Our upper bound on the probability of success provides a fairly direct operational meaning for the Hilbert-Schmidt inner product $\tr(\rho\sigma)$.

The main open problem is to prove our conjecture that every pair of states has a pair of unbiased decompositions. A notable special case of that conjecture would be when the states commute. In the other direction, a lower bound on $\Delta$ when restricted to decompositions into pure states would be more interesting than the trivial lower bound we give for the general case. Finally, it is likely that the connection with preparation contextuality can be extended beyond the very special case we consider.

\begin{acknowledgments}
  We thank K. Audenaert, J. Barrett, F. G. S. L. Brand\~ao, S. Castiglione, G. McConnell and A. Scott for discussions. Both authors are supported by the EPSRC.
\end{acknowledgments}

\bibliography{lost}

\end{document}